\documentclass[journal=jacsat,manuscript=article]{achemso}

\usepackage{graphicx}
\usepackage{dcolumn}
\usepackage{bm}
\usepackage[latin1]{inputenc}
\usepackage{mciteplus}
\makeatletter 
\def\@citess#1{\textsuperscript{[#1]}}
\makeatother 
\usepackage{achemso}
\setkeys{acs}{usetitle = true}

\graphicspath{{./eps/}{./pdf/}{./input/}{./figures/}}

\author{Marcel Reutzel}
\affiliation{Fachbereich Physik und Zentrum f{\"u}r
Materialwissenschaften, Philipps-Universit{\"a}t, D-35032 Marburg,
Germany}
\author{Niels M\"unster}
\affiliation{Fachbereich Chemie, Philipps-Universit{\"a}t, D-35032 Marburg,
Germany}
\author{Marcus A. Lipponer}
\affiliation{Fachbereich Physik und Zentrum f{\"u}r
Materialwissenschaften, Philipps-Universit{\"a}t, D-35032 Marburg,
Germany}
\author{Christian L\"anger}
\affiliation{Institut f{\"u}r Angewandte Physik,
Justus-Liebig-Universit{\"a}t Giessen, D-35392 Giessen, Germany}
\author{Ulrich H\"ofer}
\affiliation{Fachbereich Physik und Zentrum f{\"u}r
Materialwissenschaften, Philipps-Universit{\"a}t, D-35032 Marburg,
Germany}
\author{Ulrich Koert}
\affiliation{Fachbereich Chemie, Philipps-Universit{\"a}t, D-35032 Marburg,
Germany}
\email{koert@chemie.uni-marburg.de} \phone{+0049 (0)6421 2826970}
\author{Michael D\"urr}
\affiliation{Institut f{\"u}r Angewandte Physik,
Justus-Liebig-Universit{\"a}t Giessen, D-35392 Giessen, Germany}
\email{michael.duerr@ap.physik.uni-giessen.de} \phone{+0049 (0)641
9933490}

\title{Chemoselective Reactivity of Bifunctional Cyclooctynes on Si(001)}

\begin{document}

\clearpage

\begin{abstract}

Controlled organic functionalization of silicon surfaces as integral part of semiconductor technology offers new perspectives for a wide range of applications. The high reactivity of the silicon dangling bonds, however, presents a major hindrance for the first basic reaction step of such a functionalization, i.e., the chemoselective attachment of bifunctional organic molecules on the pristine silicon surface. We overcome this problem by employing cyclooctyne as the major building block of our strategy. Functionalized cyclooctynes are shown to react on Si(001) selectively via the strained cyclooctyne triple bond while leaving the side groups intact. The achieved selectivity originates from the distinctly different adsorption dynamics of the separate functionalities: A direct adsorption pathway is demonstrated for cyclooctyne as opposed to the vast majority of other organic functional groups. The latter ones react on Si(001) via a metastable intermediate which makes them effectively unreactive in competition with the direct pathway of cyclooctyne's strained triple bond.

\end{abstract}

\begin{figure}
\begin{center}
\includegraphics[width=0.4\columnwidth]{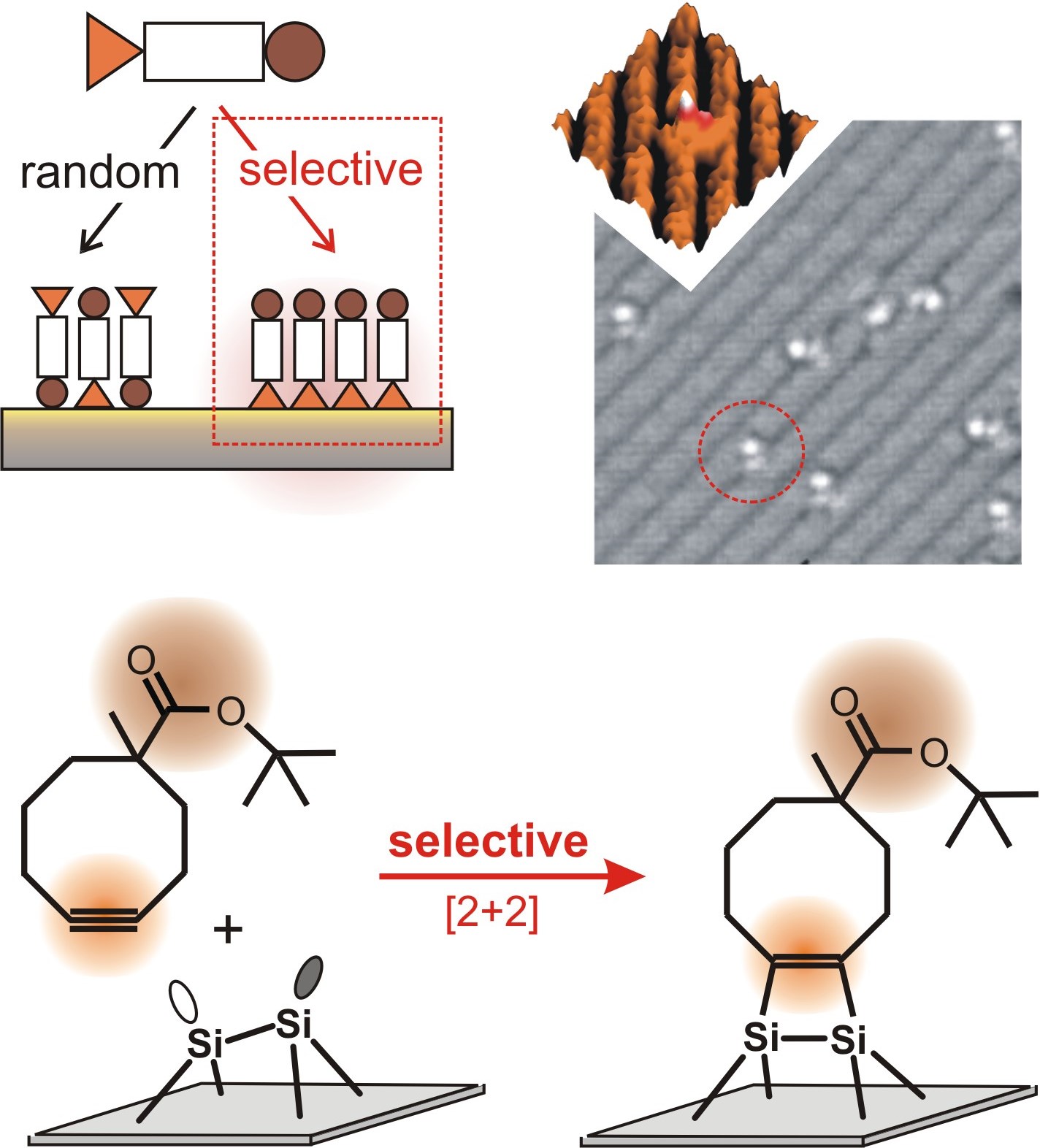}
\caption*{TOC Graphics}
\end{center}
\vspace{-5mm}
\end{figure}

\maketitle

\hyphenation{tempera-ture}


\clearpage

\section{Introduction}

Organic functionalization of semiconductor surfaces and its potential for a multitude of technological applications \cite{Yates98sci,Miozzo10jmc,Zhou13jvst} has motivated detailed investigations of the adsorption of organic molecules on semiconductors, especially on silicon \cite{Leftwich08ssr,TaoBOOK,Smeu09jacs,Wang10acr,Kachian10acr,Lim11natchem,Burliak14ChemMat,Fabre16ChemRev,Akagi16ss}.
On passivated silicon surfaces, functionalization has been successfully demonstrated via solution-based processes \cite{Burliak14ChemMat,Fabre16ChemRev}.
However, controlled organic functionalization of the pristine semiconductor surfaces, as required for a multitude of technological processes, has not been achieved so far. The challenge is the very first reaction step, i.e., the chemoselective covalent fixation of hetero-bifunctional molecules on these substrates, leaving the remaining second functional group available for further building up of complex molecular architectures, e.g., using organic click chemistry \cite{Kolb01anchem}.  The main difficulty in achieving the required chemical selectivity, especially on the technologically most relevant Si(001) surface, arises from the high reactivity of the surface dangling bonds. Due to this high reactivity, each functional group ''F`` of a bifunctional molecule adsorbs with initial sticking coefficient $s_0$ close to unity and thus the final adsorption product will typically consist of a mixture of molecules adsorbed via F1 or F2 on the surface (Fig.~1a).


With their high ring strain of 0.8~eV~\cite{Turner72}, cyclooctyne derivatives have been proven extremely useful for performing selective chemistry in vivo (''bio-orthogonal chemistry``, \cite{Sletten09anchem}).
Here we exploit functionalized cyclooctynes for selective chemistry on semiconductor surfaces (''surface-orthogonal chemistry``).
On clean Si(001), we show that cyclooctyne derivatives exclusively react with the strained triple bond in formation of 1,2-disilacyclobutene [2+2] cycloadducts, leaving the second functional group F2 untouched (Fig.~1b).
This chemoselectivity, which is observed despite the high reactivity of both functional groups, F1 and F2, on silicon, originates from the different adsorption dynamics of the reaction channels involved. In particular, we show that a direct reaction channel is operative for the strained triple bond of cyclooctyne, in contrast to adsorption via a metastable intermediate as observed for the majority of organic functional groups Si(001). The results demonstrate how chemical selectivity can be achieved even in a highly reactive environment when taking into account the dynamics of the respective reaction channels. In this study, cyclooctyne ether {\bf 1} and cyclooctyne ester {\bf 2} (Fig.~1c) were selected as model substrates because of the known cleavage reaction of aliphatic ethers on Si(001)\cite{Mette14cpc,Reutzel15jpcc} and the versatility of an ester with regard to functional group conversion.

\begin{figure}[t!]
\begin{center}
\includegraphics[width=0.4\columnwidth]{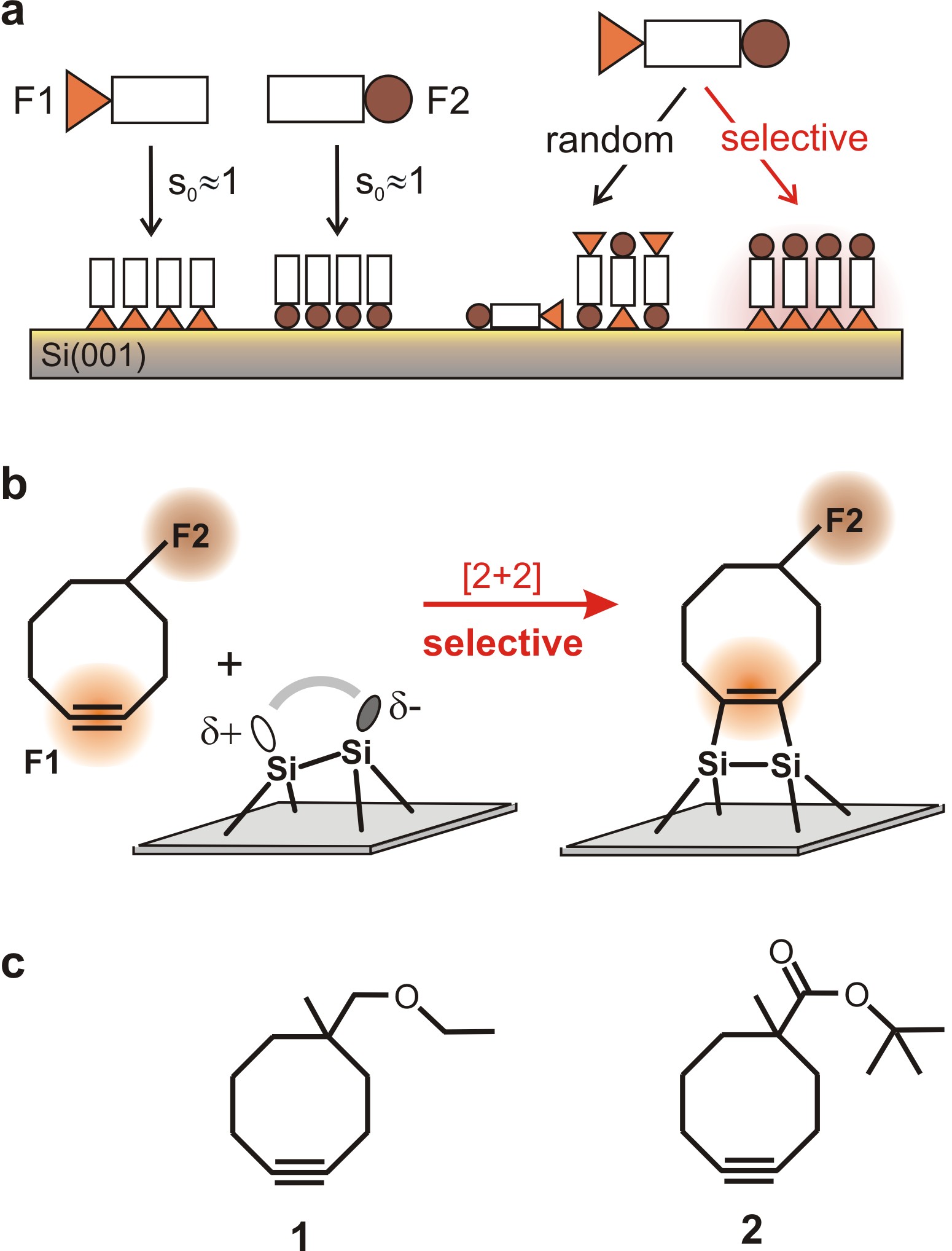}
\caption[figure1]{ a) Schematics of the adsorption process of organic molecules with one or two
functional groups F1, F2 on Si(001). The high reactivity of each molecular functionality typically leads
to random adsorption of bifunctional molecules via F1 or F2. b) Chemoselective reaction of a
bifunctional cyclooctyne on Si(001).
c) Employed test molecules.
}
\end{center}
\vspace{-5mm}
\end{figure}


\section{Experimental Section}

The XPS, STM, and molecular beam experiments were each performed in
a separate ultrahigh vacuum chamber with a base pressure $< 1\times
10^{-10}~{\rm mbar}$. N-doped silicon samples (conductivity
$\rho\approx10$~$\rm \Omega cm$) were prepared by degassing at 700~K
and repeated flashing to temperatures above 1450~K using direct
current heating. Cooling rates of about 1~K/s then result in a
well-ordered Si(001)\,$2\,\times\,1$ reconstruction
\cite{Durr02prl2,Schwalb07prb}. Cyclooctyne derivatives were dosed
through a leak valve into the ultrahigh vacuum chamber.

The XPS measurements were performed using an unmonochromatized
Mg-$\rm K_{\rm \alpha}$ X-ray source (VG Microtec XR3E2) and the
photoelectrons were detected by a hemispherical electron analyzer
(Specs Phoibos 150). The energetic position of the core-level peaks
was calibrated with respect to the binding energy of the Si~2p(3/2)
peak ($99.4$~eV). The width of the peaks was fitted by a sum of
voigt profiles. The
fitting parameters are summarized in Tab.~S1. STM experiments were performed using an {\sc Omicron} variable
temperature STM.

The measurements of the normalized initial sticking coefficient
$s_0/s_{max}$ were performed in a 4-stage molecular beam
apparatus~\cite{Durr99jcp}, using the King-and-Wells
method~\cite{King72ss} for diethyl ether. In the case of cyclooctyne, both, optical second-harmonic generation~\cite{Durr06ssr} and Auger-electron spectroscopy was employed
for measuring the initial sticking probability (for more details,
see SI).


\section{Results}

We first examine adsorption of cyclooctyne ether {\bf 1} on Si(001) by means of XPS (Fig.~2).
 The comparison of the O~1s core level spectra of cyclooctyne ether {\bf 1} on Si(001)
with those of diethyl ether ($\rm Et_2O$) on Si(001) clearly excludes a chemical interaction of
the oxygen atom of the ether group with the Si surface (Fig.~2a,b).
 The O~1s binding energy of 533.4~eV (Fig.~2a) neither matches that of the
dative Si-O bond observed for $\rm Et_2O$/Si(001) at low temperature (Fig.~2b,
blue) nor that of the covalent Si-O bond observed at room temperature (Fig.~2b,
red).
 In contrast to $\rm Et_2O$/Si(001), the O~1s peak measured for cyclooctyne ether {\bf 1} on Si(001)
does not shift in the temperature range 80--300~K and its width is typical for
oxygen in a single chemical environment (details on data and fitting are found in the Supporting Information). It is
thus attributed to oxygen atoms in a C-\textbf{O}-C environment of the still intact
ether group.
 The C~1s spectra are composed of three peaks.
 Going from high to low binding energies, they are assigned to carbon atoms in
C-\textbf{C}-O, C-\textbf{C}-C, and C-\textbf{C}-Si environments, respectively
\cite{Reutzel15jpcc,Hovis97jpcb}; compare also the C~1s
spectrum of $\rm Et_2O$/Si(001) (Fig.~2b, red), which is composed of these three
components as well.
 The intensity ratio of
C-\textbf{C}-O~:~C-\textbf{C}-C~:~C-\textbf{C}-Si measured at 300~K
matches well the expected ratio of cyclooctyne ether {\bf 1} adsorbed via [2+2]
cycloaddition of its triple bond, i.e., 2~:~8~:~2, with two carbon atoms bound to
silicon and two carbon atoms in the intact ether group (compare sketch
in Fig.~2a).
 Both, the O~1s and the C~1s spectra thus show exclusive adsorption of
cyclooctyne ether {\bf 1} on Si(001) via the strained triple bond.
This observation holds up to one monolayer coverage (compare Fig.~S1, Supporting Information).

\begin{figure*}[t!]
\begin{center}
\includegraphics[width=1\columnwidth]{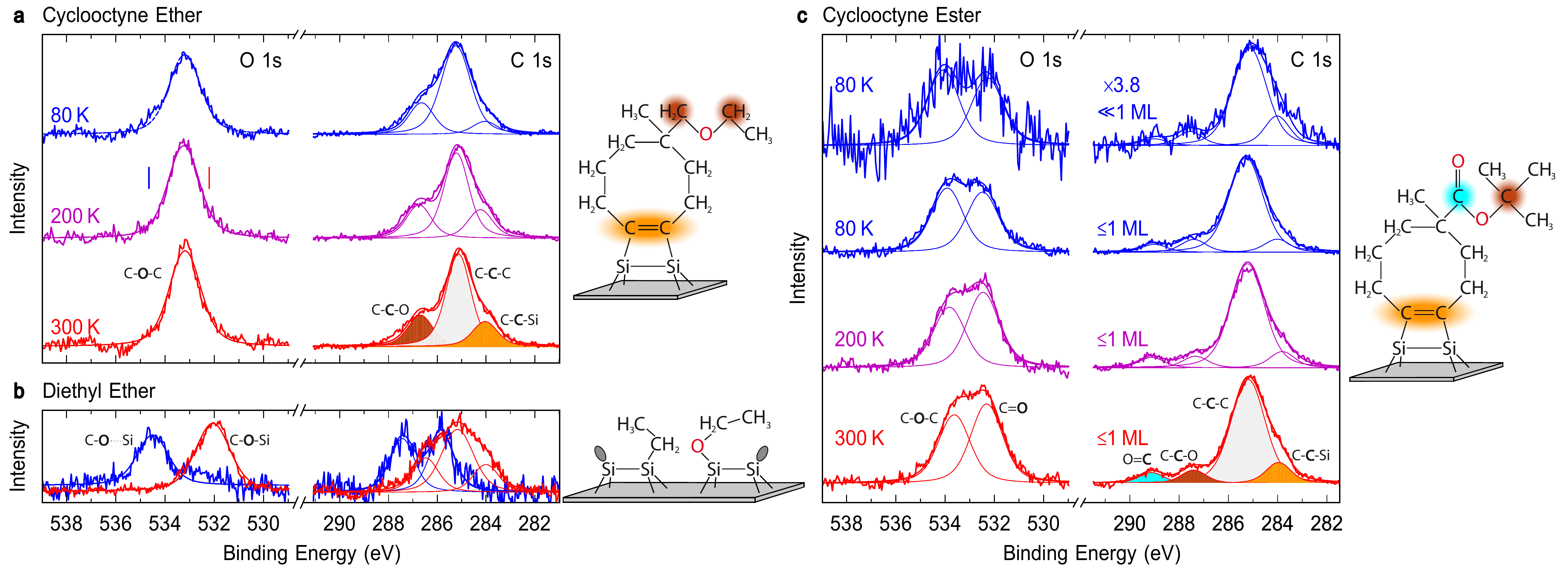}
\caption[figure2]{a) Oxygen and carbon 1s spectra of a submonolayer of cyclooctyne ether~{\bf 1}; b) of diethyl ether;  c) of cyclooctyne ester~{\bf 2}.
 All spectra were measured at 80~K surface temperature after dosing at 80~K and
annealing to the indicated temperature.
 Thin solid lines indicate the components as obtained by fitting Voigt
profiles to the spectra; parameters are provided in Table~S1, Supporting Information.
 The C~1s intensity ratios,
C-\textbf{C}-O~:~C-\textbf{C}-C~:~C-\textbf{C}-Si~=~2~:~8.8~:~1.6 (in (a), 300~K)
and
\textbf{C}=O~:~C-\textbf{C}-O~:~C-\textbf{C}-C~:~C-\textbf{C}-Si~=~0.8~:~1~:~10.3~:~1.6
(in (c), 300~K) closely match those of the different carbon chemical
environments of the adsorption geometries sketched on the right side of the
spectra. Et$_2$O/Si(001) data taken from Ref.~\citenum{Reutzel15jpcc}.}
\end{center}
\vspace{-5mm}
\end{figure*}

 Like for cyclooctyne ether {\bf 1}, the XPS spectra
of our 2nd test system, cyclooctyne ester {\bf 2} on Si(001), do not exhibit
significant changes as a function of exposure or temperature indicating a molecule in its final adsorption state (Fig.~2c).
 In particular, the O~1s and C~1s spectra show that the ester group does not react with Si(001) when it
is part of the derivatized cyclooctyne {\bf 2}: On one hand, the O~1s spectrum consists of two components, corresponding to the
two inequivalent oxygen atoms of the unreacted ester group (533.8~eV: C-\textbf{O}-C, similar to the ether group in Fig.~2a; 532.5~eV: C=\textbf{O}~\cite{Hwang04jpcb}).
On the other hand, the C~1s spectrum consists of 4 peaks of which the high-binding-energy component (289.3~eV) is assigned to \textbf{C}=O \cite{Hwang04jpcb,Lee08prb}
and thus excludes a
C-\textbf{O}-Si environment as the origin of the O~1s peak at 532.5~eV.
The observed intensity ratio of the four components of the C~1s
 spectra agrees well with the ratio
expected for adsorption via the C-C triple bond (Fig.~2c).


\begin{figure}[b!]
\begin{center}
\includegraphics[width=0.45\columnwidth]{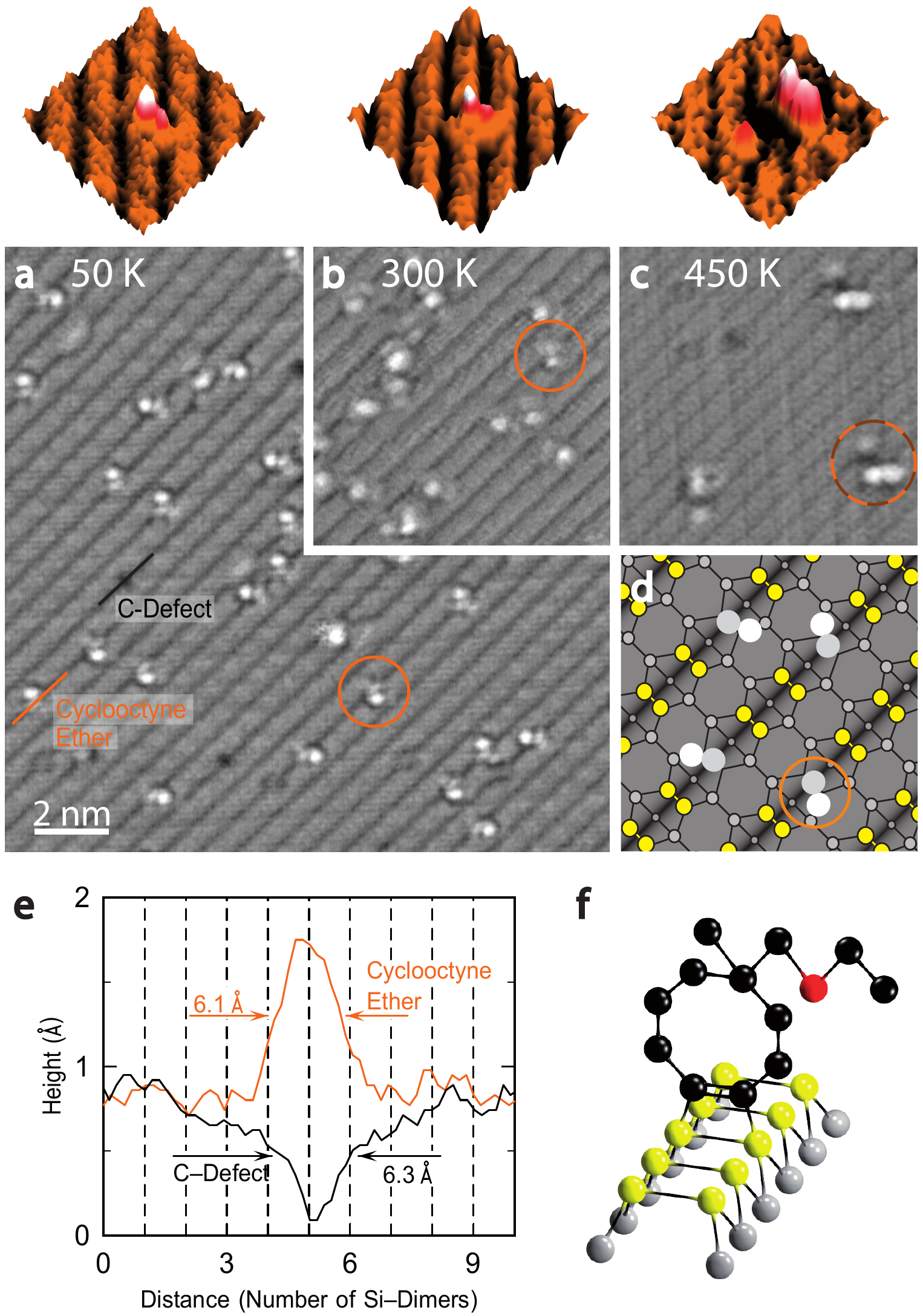}
\caption[figure3]{a) to c): Empty state STM images ($U_{\rm tun}$ = 1.9~V, $I_{\rm tun}=1$~nA, 50~K) of cyclooctyne ether {\bf 1}
adsorbed on the Si(001) surface. The nodal plane between the dangling bonds of
the Si dimers appears black and separates the Si dimers of the dimer rows as sketched in (d). In (a), 0.05 monolayer (ML,
1~ML~$\hat{=}$~1~molecule per silicon dimer) were adsorbed at 50~K
(image size 16$\times$16~$\rm nm^2$); for (b), the surface was
annealed at 300~K (0.06~ML, 9$\times$9~$\rm nm^2$). In (c), the surface is shown after annealing to
450~K (0.01~ML, 9$\times$9~$\rm nm^2$). On top of (a) to (c), pseudo-3D STM images of the respective configurations are shown. e) Height profile along the center line of the dimer row across a cyclooctyne ether {\bf 1} feature (orange) as well as across a missing dimer defect (''C-defect``, black). Both features show a similar extension along the dimer row which is substantially lower than the width of two dimers, i.e., 7.6~\AA . f) Sketch of the adsorption configuration.}
\end{center}
\vspace{-5mm}
\end{figure}

STM experiments on the adsorption of cyclooctyne ether {\bf 1} on Si(001) were performed at
low coverage to complement the XPS results on the single adsorbate level
(Fig.~3).
 Both, in images taken directly after adsorption at 50~K (Fig.~3a) as well as
after annealing at 300~K (Fig.~3b), a single, asymmetric adsorption
configuration is observed; it is located on one dimer row and consists of two bright spots of different intensity.
 If the surface is further annealed to 450~K, a different situation arises: The
bright features are still resolved (Fig.~3c), but the configurations are more
complex and extend across two dimer rows with an additional dark and bright
spot next to the original bright features. Such adsorption configurations
extending over two dimer rows are typical for ether cleavage on
Si(001)~\cite{Reutzel15jpcc,Mette14cpc}.
 At 450~K, we thus observe adsorption with both functionalities of cyclooctyne
ether.
 As such configurations extending over two dimer rows do not appear at
50~K and 300~K, the chemoselective reaction of cyclooctyne ether {\bf 1} via the triple
bond is also operative at very low coverage.
 By inspection of the lateral extension of the 50~K and 300~K adsorption
features (Fig.~3e), especially in comparison with defects/adsorbates extending over one and two dimers along the dimer row (see
also Supporting Information, Fig.~S2), adsorption on-top of one silicon dimer can
be concluded (Fig.~3f), in agreement with earlier results on the adsorption of
non-derivatized cyclooctyne \cite{Mette13cpl}. The latter shows adsorption in the on-top configuration up to one monolayer coverage.


\begin{figure}[t!]
\begin{center}
\includegraphics[width=0.45\columnwidth]{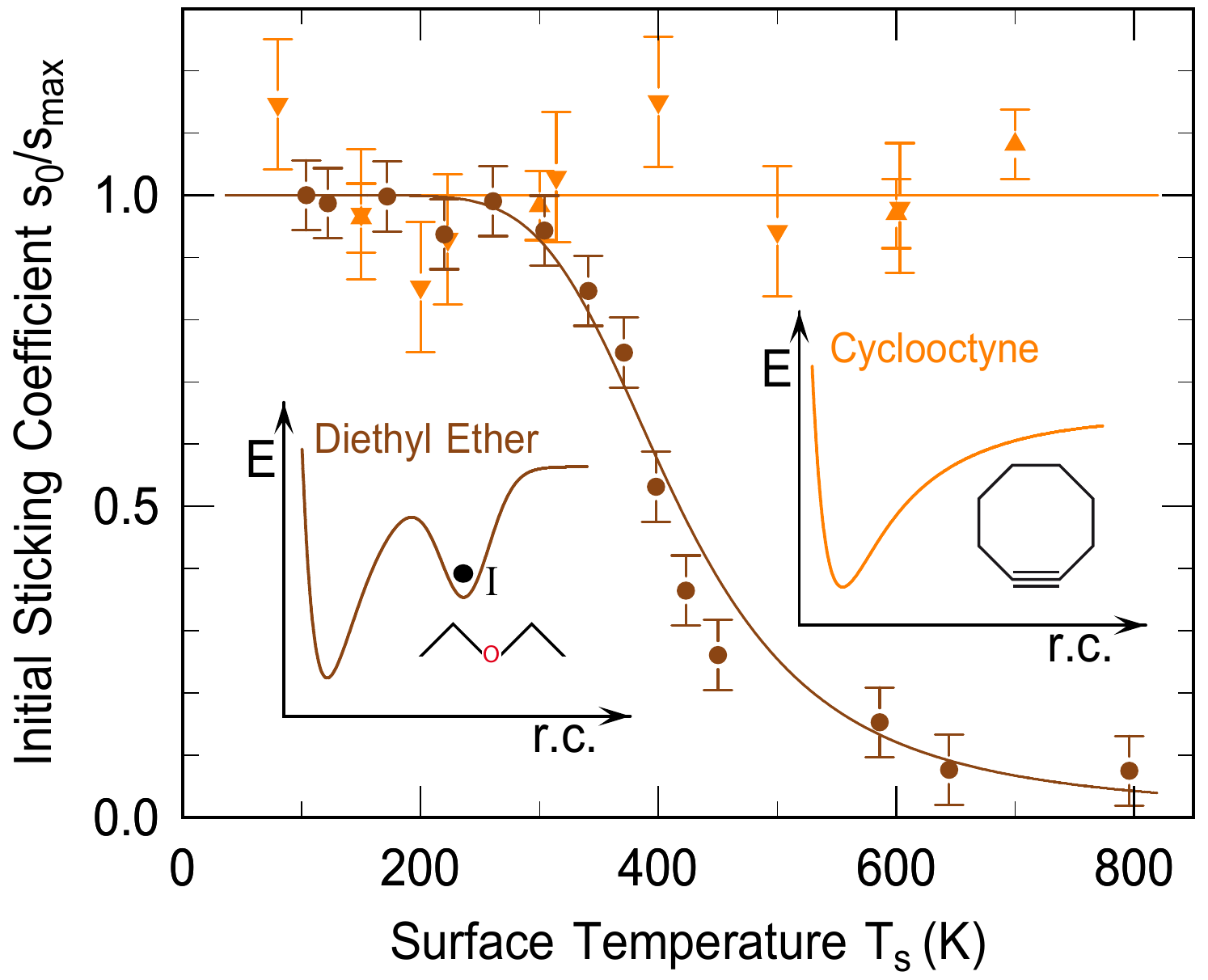}
\caption[figure4]{Normalized initial sticking coefficients $s_0$ on Si(001) as a function of surface temperature for cyclooctyne (orange triangles) and diethyl ether (brown dots). Left inset: schematic potential energy diagram for adsorption via an intermediate state I (Et$_2$O, quantitative description by the Kisliuk precursor model \cite{Lipponer12jcp}, brown line in main panel); right inset: schematic potential energy diagram for a direct adsorption pathway (cyclooctyne). E$_2$O data taken from Ref.~\citenum{Reutzel15jpcl}.}

\end{center}
\vspace{-5mm}
\end{figure}

The XPS and STM measurements unambiguously show the chemoselective adsorption of cyclooctyne ether/ester via its triple bond. In order to gain mechanistic insight, initial sticking probabilities $s_0$ were measured as a function of surface temperature for the separate functionalities, i.e., cyclooctyne and diethyl ether, on Si(001) (Fig.~4). In the case of Et$_2$O/Si(001), $s_0$ stays constant for moderate surface temperatures  $T_s$ but decreases with higher $T_s$ (Fig.~4, brown circles).
This behavior is typical for organic molecules on semiconductor surfaces \cite{Clemen92ss,Lipponer12jcp}.
Their adsorption is overall non-activated but proceeds via a kinetically stabilized intermediate state (Fig.~4, left inset).
Once trapped in this intermediate state, the adsorbate can either convert into the covalently attached final state or desorb back into the gas phase.
With rising surface temperature, the relative weight of the desorption rate increases and thus a decrease of $s_0$ is observed.
In contrast, for cyclooctyne on Si(001), $s_0$ stays constant over the whole temperature range (Fig.~4, orange triangles).
This indicates a direct reaction channel into the final state (Fig.~4, right inset), from which desorption is negligible even at highest surface temperatures.


\section{Discussion}

\begin{figure}[t!]
\begin{center}
\includegraphics[width=0.6\columnwidth]{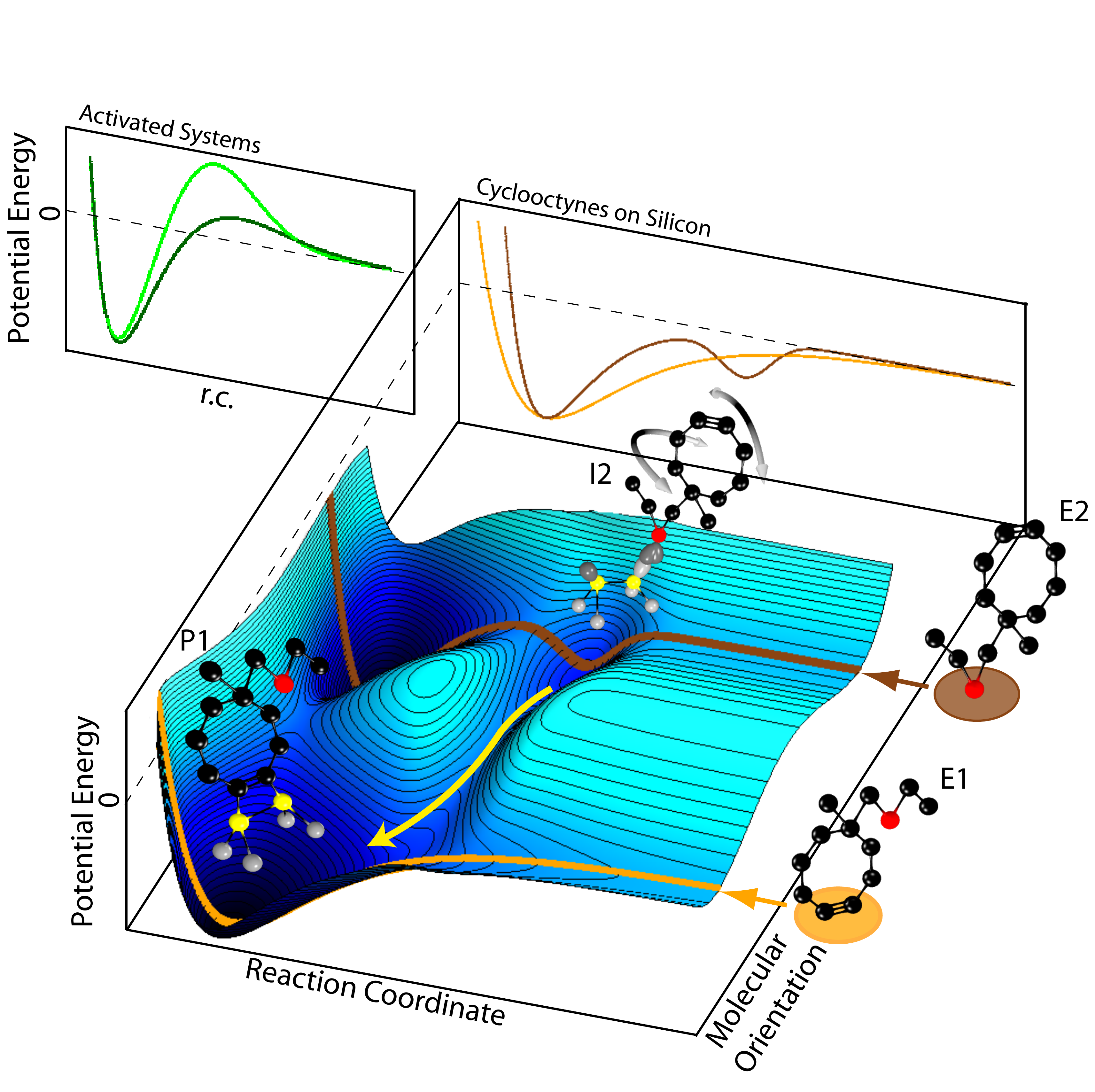}
\caption[figure5]{Main panel: Combined potential energy surface for the functional groups of cyclooctyne ether {\bf 1} (orange: strained triple bond, brown: ether group). Depending on
molecular orientation when approaching the surface (reactants E1, E2), the
molecule adsorbs either directly via [2+2] cycloaddition into the final state
(E1 to product P1) or is trapped in the datively bonded intermediate state (E2 to I2)
with subsequent [2+2] cycloaddition of the triple bond and concomitant release of the
dative Si-O bond (yellow arrow between the two channels). In the back, the projection of the two potential energy curves of the two functional groups on Si(001) are compared to chemoselectivity in the case of activated processes (green curves).
}
\end{center}
\vspace{-5mm}
\end{figure}

This direct reaction channel of the strained triple bond is the basis for the chemoselective adsorption of derivatized cyclooctynes on silicon.
 In order to illustrate the general principle, we consider the movement of cyclooctyne ether {\bf 1} on a schematic potential energy surface as depicted in Fig.~5: Molecules approaching the surface with the triple bond will react directly in the [2+2] final state
(Fig.~5, reaction of educt E1 to product P1). Molecules approaching the surface with the ether group will be transiently trapped in the intermediate state I2 (Fig.~5,
from E2 to I2).
In this weakly bound intermediate, the trapped molecule typically still has a high degree of rotational and
vibrational freedom.
 It allows the bifunctional molecule to sample the surface with the strained
triple bond during its finite lifetime in the trapped state ($\approx 100$~ms for Et$_2$O at 300~K \cite{Reutzel15jpcl}), thus enabling the [2+2] cycloaddition of the cyclooctyne ring (Fig.~5, yellow arrow).
According to our results, this [2+2] cycloaddition goes along with the release of the
dative Si-O bond, leading to a molecule solely bound via the cyclooctyne unit.
As most functional groups react on Si(001) via such a weakly bound intermediate state, our concept for chemoselectivity based on cyclooctyne as functionality F1 is largely independent on the details of F2, in distinct difference to other, more restricted approaches discussed in literature
\cite{Hossain04cpl,Shao09cpl,Ebrahimi09ss,Zhang11jpcc,Shong15jpcl}.

The observed chemoselectivity of functionalized cyclooctynes on silicon is based on the different adsorption dynamics of the two reaction channels involved, which are both overall non-activated. This is in pronounced contrast to other chemoselective systems, which are typically overall activated (Fig.~5, green lines). In the latter case, chemoselectivity arises from the difference in the energy barriers involved (Fig.~5, light versus dark green line). Examples are chemoselective adsorption on passivated semiconductor surfaces \cite{TaoBOOK,Ulman96ChemRev,Burliak14ChemMat} and chemoselective reactions in heterogeneous catalysis \cite{Ertl08AngChem}.


\section{Conclusion}

In conclusion, we introduced a general concept for chemoselective adsorption on silicon surfaces based on functionalized cyclooctyne derivatives.
We exploit the fact that most organic functionalities (F2) react on Si(001) via a weakly bound metastable intermediate state from which the
reaction of the strained triple bond (F1) is still accessible and is in fact
kinetically more favorable than the further reaction of group F2. This second functional group is then available for the attachment of further organic reagents.
As F2 is not subject to major restrictions, it can be chosen according to the reaction scheme employed.
Our results thus open the road to a selective covalent molecular architecture on semiconductor surfaces.
Beyond that, the results show in more general how chemoselectivity can be achieved for highly reactive systems when exploiting the adsorption dynamics of the respective reaction channels.




\medskip

\clearpage

\setlength{\parindent}{0pt}

{\bf Associated Content}

{\sl Supporting Information}

Details on molecular beam experiments and XPS fitting procedure;
coverage dependent XPS measurements of cyclooctyne ether {\bf 1} on
Si(001); STM height profiles on the adsorption configuration of
cyclooctyne ether {\bf 1}; details on synthetic route for synthesis
of the cyclooctyne derivatives {\bf 1} and {\bf 2} used.

\bigskip

{\bf Author information}

{\sl Corresponding authors}

koert@chemie.uni-marburg.de (U.K.);\\ michael.duerr@ap.physik.uni-giessen.de (M.D.)

\smallskip

{\sl Notes}

The authors declare no competing financial interest.

\bigskip

{\bf Acknowledgment}

Funding by the Deutsche Forschungsgemeinschaft through SFB~1083 and GRK~1782 is gratefully acknowledged.

\clearpage

\setkeys{acs}{usetitle = true}

\bibliography{cyclobib}

\end{document}